**Electrical Conductance in Biological Molecules**

By M. Waleed Shinwari*, M. Jamal Deen*,***, Evgeni B. Starikov**,**** and Gianaurelio Cuniberti**,***

[*]     M. Waleed Shinwari, Prof. M. J. Deen
Department of Electrical and Computer Engineering
McMaster University, Hamilton, ON L8S4K1  Canada
E-mail: (jamal@mcmaster.ca)

[**]     Prof. Evgeni B. Starikov, Prof. Gianaurelio Cuniberti
Institute for Material Science and Max Bergmann Center of  Biomaterials
Dresden University of Technology, D-01062 Dresden, Germany

[***]   Dept. of Computer Science and Engineering
Pohang University of Science and Technology Pohang, Korea.

[****] Department of Chemical and Biological Engineering, Chalmers University of
Technology, SE-412 96 Göteborg, Sweden.

Keywords: DNA, conductance, molecule, quantum, nano

(Nucleic acids and proteins are not only biologically important polymers: They have recently

been recognized as novel functional materials surpassing in many aspects the conventional

ones. Although Herculean efforts have been undertaken to unravel fine functioning

mechanisms of the biopolymers in question, there is still much more to be done. This

particular paper presents the topic of biomolecular charge transport, with a particular focus on

charge transfer/transport in DNA and protein molecules. Here the experimentally revealed

details, as well as the presently available theories, of charge transfer/transport along these

biopolymers are critically reviewed and analyzed. A summary of the active research in this

field is also given, along with a number of practical recommendations.)

## 1. Introduction

With the increasing diversification of applications requiring materials with specific electronic properties, and with the continuing thirst for miniaturization and packed integration of electronic devices, attention is being diverted from the optimization of regular semiconductors (Si, Ge, GaAs, and other elementary and compound semiconductors) to the investigation of newer compounds that have semiconductor behaviour, particularly organic materials. These materials, once treated as insulators, are showing promising electrical properties now that our technology is capable of detecting much lower currents and probing the molecular structures of these materials in much more detail. Applications that require the use of organic semiconductors and conductors range from biomedical equipment and sensors to home theatre and TV systems. The main advantages of using organic materials in place of conventional semiconductors are their lower current and power operation, cheaper and simpler fabrication, versatility in usage, and mechanical flexibility which allows electronic devices to be incorporated into fabrics and flexible plastic structures. The main disadvantages to using organic semiconductors are their low life time due to degradation, as well as their reactivity with water and other substances, which necessitate the design of effective packaging systems. Conventional semiconductors rely on the transport of charge carriers (electrons and holes) through the semiconductor crystal by using delocalized conduction and valence band electronic states. In organic semiconductors, due to the intricate structures of the underlying molecules, we have as a rule no states that are delocalized over the entire structure, but instead they are localized over one molecule or part of a molecule. In accordance with this, there are no conventional "valence and conductive bands", but lots of localized electron states with comparable energies. Depending on the spatial extent of these energy states, electrons occupying these states will then have a nonzero probability of "hopping" between these states. Basically, the hopping process could be considered a kind of quantum-mechanical tunnelling via the energy barrier throughout the separation between the two states. If the two states are at exactly the same energy, the probabilities of tunnelling are equal in both directions and, as a result, no net current can be expected in this case at zero bias. However, if one of these states is made to have a lower energy using some external factor, then the tunnelling probabilities will not be in balance, so that a net current should flow.

Until a few decades ago, research in organic charge transport has dealt with the bulk solid-state amorphous collection of the organic molecules. However, recently, there spiked an interest in understanding the transport of charge through single organic molecules, with the special focus cast on biological molecules such as DNA and proteins. The approaches to these molecules should be fundamentally different from those working for any other inorganic or non-bioorganic substance, because the biopolymers in question have very high flexibility and many degrees of freedom. As a consequence, their structural and dynamical behavior is much more complicated than that of artificial polymers. Moreover, they carry net electric charges and thus may generally be dissolved in electrolytic solvents. Finally, their shapes and conformations are highly dependent on the solution's concentration, pH value, temperature etc. With the electric properties in mind, it is then clear that providing reliable Ohmic contacts between single biopolymers and the corresponding electrodes can be especially challenging. This paper intends to discuss the topic of biomolecular charge transport, with a particular focus on charge transport in DNA and protein molecules. The paper starts with an assessment of the reasons for conducting such research. From there on, different experimental results, theories and details of charge transport along these molecules will be scrutinized, and finally a summary of active research in this field will be given, along with some practical recommendations.

## 2. The Need for Molecular Conduction

Why are we interested in studying the electrical conductance of biomolecules? After all, an appreciable number of commercial electric devices is based on the bulk conduction of organic polymers, and the conventional transport theories focus rather on the carrier transport between the neighboring polymer molecules and not on that within the molecule itself. Processes such as polaron-assisted tunnelling and phonon-assisted hopping are prime candidates in theories for charge migration between amorphous polymer segments, but the charge transport across the molecule itself is not detrimental because the electron states are delocalized within the molecule and the hopping process is what determines the overall conductivity. If that is the case, why bother studying transport processes within a molecule?

To answer this question, we need to re-assess the different fields that would benefit from this research. In studying single-molecule conductivity, we are not necessarily targeting easier fabrication of electronic devices or the development of flexible TV displays. The main beneficiary of molecular charge transport research ought to be the medical community. Most biochemical processes in the metabolism are oxidation and reduction reactions. Many of these reactions involve complex molecules such as proteins. Generally, such reactions are very unfavourable unless a catalyst enzyme is present. The enzyme can catalyze the reaction in many ways, one of which is by providing a means for charge transfer between the two molecules. Understanding electron transport in these systems should in principle provide us with a possibility to alter the enzyme or the reactants in such a way that the chemical reaction under study is accelerated or inhibited. Electronic control over metabolism could be a hot topic in medicine, particularly in cancer therapy.

Another medical use of this research is to design better optimized biosensors. Most current DNA and protein biosensors suffer from large variances that are either due to random external noise or owing to ensemble averaging of a large amount of the analyte. By understanding single-molecule electron dynamics and mastering the fabrication of single-molecule devices, we can design robust single-molecule sensors that are very specific to the sensed analyte, and we can also enhance the sensitivity of current conductance-based biosensors.

Aside from the medical domain, several research areas can also benefit from biomolecular conduction. When working with molecular electronics and nano-electromechanical systems (NEMS), it becomes extremely hard to form stable nanostructures using crystalline or polycrystalline materials, such as done now in NEMS. At such molecular scales, one needs to search for molecules with particular shapes and conformations. Biology is rich with molecules of varying shapes, and organic chemistry is the chemistry of shape and conformation. With organic elements, we can synthesize molecules possessed of specific shapes to perform specific functions. Actuating these nanomachines requires current transport through the molecule. It is therefore extremely important to understand the conductive behaviour of such molecules and optimize the design of the actuators.

Besides, in molecular electronics, there is a continued demand for conductive nanowires to act as interconnections between circuit elements. Linear polymers such as DNA provide such molecules, and the study of their conductive properties is therefore of high importance in designing molecular electronic devices. The self-organizing and specific-binding properties of DNA molecules will allow the design of self-forming circuits. Finally, designing state-of-the-art optical devices requires semiconductors with carefully engineered band gaps and sharp emission and absorption spectra. Organic molecules can provide such materials due to the wide range of various electronic states with different energies that depend on the mutual conformation of the atoms. The conductive and radiative properties of charge carriers within the molecule are of crucial importance for deducing the color and intensity of the emitted light.

### 3. The Structure of Biomolecules

DNA molecules are the linear biopolymers that encode the footprints of all known organisms. A DNA molecule belonging to a certain organism encodes all of the information needed to build every cell, tissue, and membrane of the organism. It does so by carrying the information serially in the different monomer sequences along the DNA chain. A contiguous selection of these monomers that is responsible for a certain trait of the organism is known as a gene. The molecular geometry of DNA is covered in detail in many textbooks (see, for example, [1]), so we wouldn't like to dwell on this topic here. Similar to the DNA molecule, proteins are essentially one-dimensional polymers. The basic building block of a protein is called an amino acid, and it always contains an amino group (NH3) and a carboxyl group (COOH-), in addition to some special organic group (side chain) which distinguishes different kinds of amino acids. We send the interested readers to any biochemistry textbook for the protein structure details.

## 4. Conductance Experiments in Biomolecules

The electrical conductivity of biological polymers is of extreme importance in characterizing a lot of biochemical reactions and in determining the usability of biomolecules in sensors and nanoelectronic circuitry. Before studying the physics of the conductivity in the biomolecules, we need to measure and verify the conductivity of these molecules. The main problem here is: how do we attach electrodes to single molecules? Would such a connection not compromise the current-voltage characteristics of the device by incorporating its own current-voltage characteristics in series with the molecules of interest? How do we verify that any recorded conductivity comes from the molecule itself (DNA or protein) and not from some residual conductivity in the surrounding medium?

Therefore, the first step in establishing good experimental results is to provide reliable electrical contacts to a single molecule. These electrical contacts should not allow any electron transfer reactions through the ionic medium surrounding the molecule, or at least the electrical conduction by other means should be carefully characterized and accounted for using control experiments. While presenting a thorough review on the biopolymer conduction experiments is not our intent – because of so many excellent and comprehensive treatises in the field (see, for example, [2], [3], and the references therein) – here we would solely like to pick out a number of characteristic examples of such studies to demonstrate merits and demerits of the modern approaches.

## 4.1. DNA Conductance Experiments

Some of the earlier work on the conductance of single DNA molecules has delivered contradicting results, ranging from metallic to insulating behaviour [4-7]. Experiments were performed on single DNA molecules suspended between electrodes or on bundles of DNA, and under various conditions. The discrepancy in the results is a clear indication of the complexity of the experimental setup and the presence of many interfering phenomena. Among the first groups to report semiconductive behaviour of DNA and provide a measure of its electrical bandgap was Porath et al [8]. A DNA molecule was electrically trapped on a platinum-coated broken $SiO_2$ bridge, as shown in **(Figure 1)**, forming a connection between the two ends. The experiment was conducted in vacuo to eliminate any residual conductance from the surrounding medium. Although the quality of the electrical contact had been questionable, this experiment provided the basis for further investigations on the semiconductive nature of DNA.

In the experiment of Porath et al, the possibility of the Coulomb blockade effect due to poor contacts cannot be dismissed, and might contribute to the observed voltage gap. Additionally, the edges of the DNA molecule were lying horizontally on the platinum electrode. Later it has been shown [9] that a tight contact of the DNA with a substrate can cause significant deformations within the DNA. As such, it is important to try establishing a strong direct contact to the DNA with minimum overlap of the electrode. In addition, the electrode material of choice and its Fermi level must be close to that of the DNA for good contact establishment. One attempt at addressing these issues was the work of Watanabe et al [10], which involved the use of carbon nanotubes (CNT) as electrodes in contact with DNA. Although the proper covalent contact was not established, the measurements at room temperatures of the conductance of the DNA-CNT assembly (shown in Figure 2) showed a voltage gap of 3V at room temperature. Following these early pioneering experiments, several other experiments attempted to further probe the conductance of DNA while trying to circumvent different sources of contamination. Xu et al [11] carried out conductance measurements of DNA in a salt buffer to allow it to maintain its native conformation. The DNA terminal was thiolated and chemically attached to gold contacts. Interestingly, histograms of conduction showed peaking counts of conductance on integers of a fundamental "conductance quantum" value (Figure 3). This was interpreted as the successful electron transfer through an integer number of DNA bridges. A retracting force-conduction measurement also showed step-wise decrease in conductance, which was attributed to sequential break-up of DNA bridges. While such experiments provide a very unique approach of measuring DNA conductivity, they are not conclusive in determining the actual mechanisms of charge transport. Due to the small length of DNA (8-14 bp), electrons might be able to tunnel between the electrode-contacted thiol states on either end of the molecule, giving the impression of molecular conductance and even eluding control experiments, as the thiols would be absent in absence of the DNA bridge. The I-V plot did not show any conductance gaps, which could support the hypothesis of tunnelling current through the small gap or even ionic conduction in the environment.

One of the earliest attempts at measuring the DNA conductance using scanning microscopy was done by Xu et al [12], in which a bias was provided between the STM probe and a gold substrate containing thiolated bound DNA molecules in ultra-high vacuum. The STM images showed that the bias-dependent DNA conformation can affect its conductance, and that flat-lying DNA on gold can be more conductive than free-standing DNA, indicating the possibility of hybridization of the DNA electronic states with those of the gold electrode and hence conduction enhancement. In addition, I-V characteristics on standing DNA were obtained, showing that the DNA layer was successful at inhibiting conduction up to the voltage bias of 2.8 V, and then the steep onset of conduction could be observed. This finding supports the theory of wideband semiconductive behaviour of DNA. It is important to note, however, that in this experiment, the distances between the gold electrode and STM tip were very small, so that tunnelling currents could be indiscernible from the DNA conduction current, as shown by the background tunnelling conduction in absence of the DNA. Nevertheless, this experiment has shown that the DNA molecules do contribute a voltage gap and are therefore exhibit semiconductive behaviour.

To sum up, the experiments of Xu could not provide a reliable contact between the STM tip and the DNA. The work of Cohen [13] circumvented this problem by introducing gold nanoparticles (GNP) covalently bonded to the vertically placed DNA molecules. Atomic force microscopy (AFM) probes were used to contact to the GNP, as shown in (Figure 4). The probe would now provide a more reliable electronic contact and eliminate any Coulomb blockade effects. The results obtained here confirmed the voltage gap of the DNA but also predicted currents in excess of 100 nA per molecule above threshold bias, suggesting a coherent band transport within the DNA. To assess the importance of the thiol connection to the GNP, the same group compared conductance results with and without the thiol linkers [14] and found that unreliable contacts can cause a reduction in the observed current. This work also demonstrated the inefficiency of single-stranded DNA for current conduction by using the AFM probe in open-loop mode and measuring the conductance as a function of the z-displacement from the gold surface. It was shown that conductance for ssDNA probes does not start until the AFM probe was close enough for tunnelling current to flow between the two gold electrodes, whereas the dsDNA started conduction at a larger displacement.

The above-mentioned experiments, along with many others that were carried out during the same time, have shown using different nanotechnology methods that a single DNA molecule is capable of conducting current, and does show semiconductive behaviour under the appropriate conditions. These experiments have demonstrated the essential difference between the models of legacy bulk material conductors and molecular conductance, including the need for reliable contacts as well as the effects of surface proximity and environmental parameters. However, all of these experiments were conducted on short DNA segments that leave the possibility of tunnelling current contamination valid. A more recent attempt by Roy *et al* [15] has attempted to measure the conductance of a longer DNA molecule with random base sequence (taken from an influenza gene). The ends of a horizontal DNA with 80 base pairs were covalently linked to the caps of similar-sized carbon nanotubes using amino links. Using carbon nanotube caps for attachment reduces any effects the surface might have on the DNA structure. The interaction of the "lying" DNA with the horizontal substrate was also eliminated by etching a trench in the $SiO_2$ layer and leaving the DNA molecule suspended between the carbon nanotube electrodes, with a reliable covalent interaction between them. I-V measurements showed the familiar s-curve with currents up to 30pA under 1V bias. This contradicts with the currents reported by [13] by more than 3 orders of magnitude, even though the G-C content (with low ionization potential, more delocalization of electron states and thus higher electron conductivity) was higher in the sample of [15] than that of [13] (see Figure 5). This finding could be used to dismiss the coherent band transport model of DNA, and suggest a hopping model, where the loss of correlation of motion along the DNA double helix can predict a rapid current decay with increased length. However, other explanations are possible, including the heavier hole transport as opposed to electron transport, depending on work function of the electrode used. The observed voltage gap was also smaller, namely < 0.5V at ambient temperature. The discrepancy in the band gap between this and the previous experiments can lead to different conclusions. A longer DNA segment can have more π-stacking and delocalization, which would result in a larger DOS (density of states) band broadening and a smaller gap. On the other hand, the amino contacts, as opposed to the thiol contacts, might provide stronger coupling to the nanotubes such that any residual Coulomb blockade effects are eliminated, so that the gap is reduced.

A very interesting recent experiment on DNA conductance was done by Guo *et al* [16]. Here, amine-modified short DNA strands were covalently attached to the edges of a carbon nanotube gap. The DNA bridge seemed to mimic the behaviour of the nanotube used (semiconductive or metallic) under some gate voltage bias applied to the DNA backbone. This finding would support the hypothesis that the contact effects are dominant for such short DNA strands (15bp here). Furthermore, the metallic nanotube-DNA current response showed a lot of irregular features which could indicate either richness in the DOS plot of the DNA under study, or a variation in the surface chemistry and counterionic charge around the DNA at different gate biases, which would accordingly affect the electronic structure of the DNA. The experiment investigated the effects of pair mismatches and found a dramatic increase in the DNA resistance for a single mismatch in the center. Although no conclusive evidence of mismatched DNA strand hybridization was given, thermodynamic data [17] shows that the de-stabilizing effect is in the range of 4 kcal/mol, which is easily overcome by the stabilization act of the rest of the strand. Still, even this single mismatch contributes around 173 meV of destabilizing energy, rendering the DNA strand hybridization less likely. Thus, the incubation time or the target concentration would have had to be made larger before the hybridization can be assumed to have taken place, so that the possibility of lack of the proper DNA strand hybridization cannot be completely dismissed. On the other hand, if the hybridization is verified, then this experiment would be a conclusive evidence of electronic conduction and would support the hopping transport model rather than the band conduction model.

Another fascinating experiment to probe the conductance of DNA was to use a mechanically controllable break junction (MCBJ) [18]. Here, the separation between two electrodes can be changed to within a resolution of less than 1 Å. Thiolated DNA was used to form a covalent bridge between the electrodes. The measured I-V characteristics in this work did not show a gap, which could be due to a background leakage conductance or due to good electrical contacts to the electrodes. However, upon stretching the DNA molecules, several conductance jumps are noticed (see Figure 6). Kang *et al* attributed these jumps to conformational changes within the DNA, which can modify its electronic structure, thus altering the degree of charge transmission and the conductance. This provides a demonstration that the DNA molecule is in fact responsible for current conduction.

An attempt to directly probe the electronic structure of DNA was done by Shapir, *et al* [19], where the I-V curve between an STM probe and a poly-GC DNA-on-gold was recorded and related to the density of states of the DNA. The measurements were done at cryogenic temperatures to limit the thermal broadening of the states. The results showed reproducible gaps in the measured density of states, and provided further evidence for the wide-gap semiconductor structure of the DNA molecule. The authors have also shown, via computer simulations, that the counterionic charge can contribute new states within the band gap, thus significantly affecting the I-V behaviour.

From the previous experiments, it seems that the consensus is that DNA double helices themselves are in fact a wide-gap semiconductors, though the mode of charge transport through the DNA remains controversial. The semiconductive behaviour of DNA was demonstrated by a number of independent groups using different nanotechnology tools and techniques. However, several interference sources can obscure the measurement results, possibly leading to differing conclusions. Some of these interference sources are given:

1. **The electrical contact**: The type of contact established between the electrodes and the DNA molecule can strongly influence its conductive behaviour. Just like in semiconductors, it is desirable to have Ohmic contacts with as low resistance as possible. However, Schottky-type contacts may also be made, and this could dominate the response of the connection. The DNA molecule might exhibit loss of ordered conformation or the formation of DNA kinks near the contact location. These factors can severely affect the conductive behaviour. Most groups that have made reliable links used thiol or amino groups linking DNA to gold surfaces. However, much more work is necessary to properly characterize these links and the effects of their attachment on the electronic structure of DNA.

2. **Mechanical stress**: DNA molecules, and almost all organic molecules of this type, are appreciably flexible which allows them to be stretched, bent, and twisted. When speaking of conduction via localized molecular orbitals instead of delocalized bands, the conductance becomes extremely sensitive to mechanical stress, and a flexible material can go the whole way from conducting, through semiconducting and finally to an insulating behavior under different forms of mechanical stress.

3. **Host solution:** The DNA molecule is considered one of the most stable biomolecules. However, this stability is rather feeble compared to other inorganic crystals or polymers. In fact, the stability of the DNA is largely maintained by an electrolytic host solution containing dissolved salts. The ordered structure of the counterion-hydration sheath around the DNA duplex (mimicking the double-helical structure of the latter) might give rise to other conductive phenomena that do not involve the DNA molecule, resulting in artefact (leakage) conduction. On the other hand, as was demonstrated in [19], tightly bound counter-ions in the host solution can introduce electron states that are detrimental to the observed I-V plots.

4. **The electrode material:** Depending on the choice of the electrode material, electron-transfer reactions might be possible between the electrode material and the electrolyte (Faradaic currents), leading to another path of ionic conduction which would contaminate the results of the DNA experiments. The relative position of the Metal Fermi level to that of the DNA also affects the equilibrium electronic structure and the observed conductance. Simplified solid-state treatment of the equilibrium state can lead to erroneous results, and theories of colloids and liquid crystals sometimes must be borrowed to explain the observed conductance.

5. **Temperature:** The intramolecular dynamics of DNA double helices is extremely intensive under normal physiological conditions, which is why it is possible for them to replicate and perform their biological functions. For electrical conductance, however, this means that the DNA molecule ought to suffer rapid structural dislocations, so that it is in fact difficult to resolve its electronic structure. Again, the solid-state picture of linear harmonic vibrational modes (phonons) is not readily applicable to such flexible molecules, necessitating the need to revisit and re-model these interactions and their effects on conductance. We return to this topic below, when discussing theoretical works on DNA electric properties.

6. **DNA environment:** Experiments are not generally done at a fixed reference state. For example, most of the published works were performed around neutral pH, but even a small variation in the pH value can have dramatic consequences for the DNA conductance. Additionally, the length of the DNA molecule is not fixed in many of the published experiments. It is still not completely clear how long-range DNA electron transfer occurs and whether or not it contributes significantly to the current. Certain phenomena that occur in long DNA strands might be absent in short segments. Finally, the choice of the DNA base sequence can have a profound effect on the conductance, so that truly systematic experiments ought to be carried out on the consensus reference base sequences.

7. **Type of measurement:** Not all experiments were performed by measuring the DC current through a single isolated DNA wire. As a rule, chemists conduct their experiments by attaching an electron donor to the DNA molecule, exciting it by a photon absorption or some chemical reaction, and monitoring the oxidative damage that happens within the DNA strand. This often works for longer molecules, whereas direct electrical studies are often used for shorter molecules.

8. **Molecular orientation:** Single-molecule experiments are usually carried out with the DNA molecule standing vertically while tethered on one end to a substrate, or laying horizontally on a surface, or suspended between two electrodes and forming a bridge between the latter. The horizontal DNA duplexes are liable for structural and electronic deformation by surface forces, whereas the experimental set-ups employing the vertically arranged ones are more complicated to perform.

In addition to the direct electronic measurements, several other works that aim at novel DNA-inspired molecular wires can shed light on the conductive properties of DNA base pairs. Indeed, the work of Kotlyar *et al* [20] described production of a stable linear synthetic polymer consisting of Guanine tetrads held by hydrogen bonds. The wire, called a G4-DNA, consists of four strands of DNA with tetra-guanine stacks. These synthetic molecules were found to be resistant to surface deformations, and the measured polarizability of the molecules demonstrated their superiority to normal DNA for electronic conduction [21]. Another attempt at achieving high conductivity DNA was done by Rakitin *et al* [22], where the proper imino proton of the bases were replaced by $Zn^{2+}$ ions. The resulting compound showed much better I-V curves than the corresponding DNA duplex itself. Finally, a very recent work by Tanaka *et al* [23] involved stretching of a single-stranded DNA along a metallic surface and probing the density-of-states signature of the individual bases using an STM tip. Although this method only succeeded for the states of guanine, it showed unambiguously the guanine density of states, demonstrating the ability to directly reveal the electronic structure of single molecules in experiments, using STM probes.

## 4.2. Protein Conductance Experiments

The electrical conductivity of proteins is much more complicated than DNA, because proteins are far less stable than DNA and, like DNA, can be appreciably polymorphic at different environmental conditions by adopting fibrous, globular, or membrane-shaped forms. Unlike DNA duplexes, proteins are seldom, if ever, maintained as linear unwound molecules, because they can form cross-links with other protein filaments at higher temperature and form an entangled mesh. Further, the resulting unfolded molecules cross-link together, and the resulting mesh becomes chaotic, so that it is thermodynamically unfavourable to separate the molecules again. For this reason, conductivity experiments are seldom done on linear, unwound proteins, unless they are very short polypeptides.

Early studies of the conductivity of proteins targeted purely industrial applications. It was postulated in [24] that wool protein is a perfect insulator when it is not hydrated. Absorbed water molecules act like dopants that give the wool semiconductive properties. In [25], crystallized bovine haemoglobin was sandwiched between a glass and a metallic conductor, and the DC conductivity was measured directly. The prediction of semiconductive behaviour of proteins was verified, and a band gap of 2.3 eV was calculated. Many different experiments on native and denatured proteins have shown similar semiconductive results. Table 1 [25] shows some of the published activation energies of some different dry proteins. Many of these results agree on the fact that hydration increases conduction dramatically, however all of these early works performed measurements on bulk dry proteins and not on single molecules. The finding that only hydrated proteins can conduct current supports the hypothesis that water dissociates into protons within the protein at a much higher rate than in bulk, and that proton conduction is the main method of electrical conductance in proteins.

As in the case of DNA molecules, observed band gaps can be contaminated by Coulomb blockade effects due to poor contacts. Reliable contacts to polypeptides are essential for proper measurement of conduction, especially since polypeptides are much more flexible than DNA molecules. Single molecule studies are generally done for short polypeptides rather than full proteins. Unlike DNA, these studies do not dispute the fact that the polypeptide can conduct current, but instead they try to find out the exact mechanism of conduction; whether the conduction is done by protons, holes, or direct electron transfer, and then to try and speculate on what exactly is the method of carrier transport by studying the dynamics of the peptide in question. The earlier work by Weinkauf *et al* [26] showed that charge transfer processes in suspended amino acids in the gas phase is directly related to their chemical reactivity. This study was done by laser photoexcitation of specific amino acids and measurements of the radiation of their chromophore sidechains. The same group have also studied the peptide conduction of photogenerated holes through short amino acids and have concluded that hole migration is the main mode of charge transport and is dependent on the type of amino acids and on the sequence. It was shown that a repetitive amino acid structure might be necessary for hole conduction [27]. These valuable results have shown the effect of the polypeptide's structure on its conductance. However, they do not conclusively explain how the conductance modulation manifests. It is possible that the conduction takes place along the chain or along other tunnelling paths within the folded polypeptide. Both of these paths can change with different amino acid residues.

One of the attempts at direct electrical conductance was carried out by Xu *et al* [28], where short (1-3 amino acid) oligopeptides were thiolated and connected to gold surfaces, forming a bridge (Figure 7). This was a similar setup to their work on DNA [11]. However, as the peptides can be protonated at different pH values, they tested the effects of varying the pH of the solution on the conductance peaks. It was found out that, irrespective of the peptide length, higher pH values resulted in reduced conductance. This is good evidence to the importance of protons in the conduction, but also raises the question as to the mechanism of conduction:

whether it is ionic (via protons) or electronic. Additionally, the I-V measurements did not show any gap, demonstrating the absence of any semiconductor-like behaviour, which contradicted earlier experiments on bulk hydrated proteins.

Longer peptides (~5mer and longer) are generally more promising for conductance investigation due to their potential use in nanoelectronics and nanomedicine. As the peptide becomes longer, however, it adopts localized conformations (secondary structure). This conformation, unlike DNA, is very sensitive to environmental variables like pH, which is why virtually all longer peptide experiments are done in controlled physiological conditions and not in vacuum or under cryogenic temperatures. The conductance in this case is strongly dependent on this conformation. Conductance experiments are generally done either by using a metal-molecule-metal junction and performing direct DC measurements, or by using a donor-molecule-metal and monitoring the charge transfer electrochemically. The former method is not very common for proteins but rather for DNA and other organic molecules. The study of Sek *et al* [29] has used scanning tunnelling spectroscopy to probe the conductance of 14-mer α-helical peptides that were doubly thiolated and covalently attached to a gold substrate and a gold STM probe. This experiment confirmed the absence of band gaps for helical peptides and also showed evidence of asymmetry in the I-V curve of the helical polypeptide (see Figure 8). This was attributed to the resonant dipoles of the peptide bond, which give an overall "depletion region" effect. However, it is worth mentioning that the observed asymmetry might be contaminated by a residual "Schottky" effect from the contacts.

The donor-molecule-metal technique is much more common for peptides. Generally, ferrocene (Fc) is used as an electron donor at one terminal of the oligopeptide, and the charge transfer is monitored using electrochemical techniques. Earlier attempts gave very different results. Marek and Heinz-Bernhard [30] immobilized Fc-labeled oligoprolines on gold and measured the electron transfer rate using cyclic voltammetry (CV) (Figure 9). The results showed an exponentially decaying dependence on the peptide length. Later similar experiments on longer peptides reported much lower transfer rates compared to this extrapolated exponential dependence [31]. This discrepancy might be related to authors' [30] use of oligoprolines which have rigid conformations, and whose molecular structure allows for inter-molecular hydrogen bonding, which can provide a means for charge transport [32]. A later attempt by the same group [33] compared the conductance of self-assembled polypeptide films arranged in parallel and anti-parallel conformations, where the anti-parallel conformation is more stabilized due to dipole-dipole interaction. This work showed that the anti-parallel conformation was less capable of conducting current and was viewed as evidence to the importance of molecular dynamics in the electron transfer process. It is important to note, however, that the observed resistance increase might be due to the effects of the interacting dipoles on the electronic structure of the peptides, or due to restricted ionic migration in the solution.

Most of the recent experiments on protein conductance are geared towards investigation of the charge transfer mechanisms. As the available body of works show, the two main candidates for electron transport in peptides are long range tunnelling (superexchange) and hopping transport, like in DNA. There is a wealth of experimental evidence to support the presence of these both processes. In [34], the electron transfer rate was monitored in peptides of different lengths using a donor-peptide-acceptor setup which were activated by radiolysis. The results showed an initial exponential decay of the transfer rate, consistent with the tunnelling hypothesis, followed by power-law decay for longer peptides, corresponding to hopping transport. This finding was shared by many researchers using different measurement techniques, including a recent work [35] where a Fc-molecule-gold setup was used to monitor the rate constants for different lengths of helical peptides.

Despite the consensus on the presence of hopping transport, the locations of the hopping sites and the hopping mechanisms are still matters of current debate. The initial hopping sites were postulated to be the peptide bonds. However, several other possibilities have been postulated. Using a structured peptide rich in aromatic side chains, Cordes *et al* [36] carried out controlled donor-peptide-acceptor experiments that showed that the aromatic side chains can lead to significant increase in current, supporting the idea of side-chain electron hopping. This was later modified to include other amino acid residues, some of which were not expected to facilitate hopping sites [37]. This led to the conclusion of a "group" effect of electron hopping, which could include the residue, neighbouring water molecules, and the adjacent peptide bond. The role of amino-acid residues in electron transport can enable some proteins to playing their electroactive biological role. However, by sandwiching completely folded protein molecules between two electrodes, Ron *et al* [38] has recently found that even proteins with no known electron transport functionality show traces of conductance (few nA per volt), whereas electrically active molecules can have much higher conductances. This was attributed to the presence of different conducting pathways in folded proteins, but it also demonstrates the effect of the peptide backbone as a possible mediating site for charge hopping.

The above experiments, in addition to many others, illustrate some of the difficulties with measuring charge transport in proteins. What renders the determination of the electrical properties of proteins very complicated is that proteins are not rigid wire-like molecules, but have different structural conformations that can allow multiple electron bridges and conductive pathways to form. Even in short oligopeptides, the conductance depends on the nature of the specific amino acids and the side chains, and their mutual orientation. There are 20 known amino acids in nature, as opposed to only 4 bases in DNA. These amino acids have a diversity of chemical behaviours that allows different protein conformations to be adopted under the proper conditions. Various protein conformations can result in many different means of electron transfer, and calculating or estimating the overall conductivity becomes extremely difficult and almost intractable. For these reasons, studies of electron transport in peptides are generally meant to facilitate therapeutic purposes and to gain deeper understanding of metabolic activities, while rarely contributing to the field of molecular electronics, despite the fact that proteins can have recognition and self-assembly properties that surpass those of DNA, making them better candidates for self-assembled molecular electronics.

## 5. Conductance Theories for Biomolecules

The observed conductance behaviour of biomolecules, especially DNA, hints at their possible use as molecular wires. Before this is efficiently done, however, we must know the dynamics of charge transfer within the molecule. Delocalized coherent transport is dismissed in biomolecules due to their lack of periodicity, random fluctuations, and limited conductance values from experiments. Even for a poly-guanine structure (prime candidate for a conduction pathway building block, owing to its relatively low ionization potential), simulations have shown that their stacking in the DNA duplex should not yield any extended states, or that the band gap is very large (For a thorough review of earlier attempts at explaining charge transport, see [3] and the references therein). In contrast to the familiar phonon-assisted hopping of solid-state physics, Schlag *et al* [39] proposed a "rest and fire" model for charge hopping along a loose peptide backbone, implemented using classical molecular dynamics simulations and chemical kinetics theory of electron transfer. This model was later used to identify critical conformations of dipeptides for charge hopping across residues [40], but it may also be very useful to rationalize the DNA charge transfer/transport mechanisms.

In determining the electronic properties of biomolecules, the molecular structure must be calculated, and a suitable transport theory must be used to describe the time evolution of the appropriate charge propagation. It must be kept in mind, however, that this structure is not static, as can be approximated in solid-state physics, but may experience dramatic nonlinear deformations. Solid-state theory is therefore not capable of capturing all aspects of the charge transfer dynamics in biomolecules. To this end, there are two commonly used methods of theoretically investigating charge transport: Molecular dynamics simulations, in conjunction with sequential chemical kinetic equations (such as Marcus-Hush theory) for electron transfer between nearest neighbours, and/or ab-initio quantum mechanical simulations on relaxed structures, coupled with a quantum electron transport model. The former if often used by chemists for protein conductance modelling, whereas the latter is used by physicists for more stable structures, like DNA.

The main candidates for the dynamics of charge migration in biomolecules are hopping transport and superexchange transport (chain-mediated tunnelling). Tunnelling transport predicts a transfer rate $k_{CT}$ that decays exponentially with the length of the molecule $R$ (or the number of chain units $N$) [41]:

$$k_{CT} \propto e^{-\beta R} \propto e^{-\beta N} ,$$  (1)

where $\beta$ is the distance decay parameter. Hopping transport is instead characterized by a weaker distance dependence:

$$k_{CT} \propto k N^{-\eta} ,$$  (2)

Evidence for the both of these processes exists in the literature for proteins and DNA [30, 35, 36, 42-45]. The general consensus is on the presence of both of these processes: superexchange tunnelling for short-range transfer, and hopping for long-range transport. However, the mechanisms of tunnelling and hopping are not yet fully understood. Several factors can influence the charge migration process:

1. **Carrier type**: Hole transport, electron transport, and even polaron transport, were all suggested as viable candidates for charge migration in biomolecules [45, 46]. Hole transfer is usually achieved by photo-oxidation of a donor group attached to a terminus of the molecule, whereas electron transfer is achieved by chemical reduction of an acceptor group, or direct electrical reduction of a terminus group by applying voltage onto a connected electrode. The structures and positions of the molecules' frontier orbitals (HOMO, the highest occupied molecular orbital, and LUMO, the lowest unoccupied molecular orbital), as well as the ionization potentials and electron affinities, respectively, of the mediating groups are essential in determining the charge transfer rate for all the carrier species. In the fascinating experiment of Elias *et al* [47], the hole and electron transfer rates were compared using synthetic electron and hole traps placed on the two chains of a DNA molecule, and an oxidizible/reducible complex. It was shown that both hole and electron migrations are significant and that hole migration has an edge over electron migration for poly(A-T) sequences. It is not clear, however, whether this efficient charge transfer is due to the tunnelling or hopping mechanism, as the distance ranges of the experiment were well within the superexchange regime.

Proteins are more chemically active and flexible molecules than DNA. It has been postulated [48] and later demonstrated [49] using ab-initio calculations that protons can form complexes with the peptide backbone and migrate along the protein, giving the possibility of another source of charge transport. However, this was shown to be conformation-dependent, and α-helical peptides turned out to inhibit such means of charge transfer. Such a conduction pathway can be important during electrical measurements, depending on the electrode's chemical activity. The protonation phenomenon can also be used to explain the observed dependence of peptide conductance on the solution's pH [50], although the pH value can also lead to conformational changes in the peptide, possibly inhibiting proton transfer by means of steric hindrance.

2. **Environmental Conditions and Molecular Conformation**: Using quantum-mechanical simulations, it was shown by Adessi *et al* [51] that structural deformations within the DNA molecules, together with the counter-ion species surrounding them, can modulate the transmission spectrum, significantly suppressing or enhancing conduction at different energies. The effect of counter-ions was later verified using a DFT model [19]. Environmental effects, including temperature, ionic strength, and pH, can all affect the conformation of the biomolecule, significantly altering its electronic structure. From the solid-state theory, it is known that rapid oscillations can lead to Anderson localization, and thus significant reduction of conductance. On the other hand, using a combination of molecular mechanics (for conformation ensemble generation) and quantum mechanics (for electronic structure calculation), Starikov *et al* [52] have shown that molecular motion and mismatches can actually increase the conductance under certain conditions and provide more coupled electronic states. In yet another work [53], quantum-mechanical simulations on a simple DNA model in contact with a heat bath showed significant electronic structure modulation, which could be useful in describing observed Ohmic behaviour in short DNA wires using a tunnelling approach. More recently, a complete ab-initio study using the non-equilibrium Green's function approach, was carried out by Song, *et al* [54], to study the effects of the correlated stretching/twisting of a (GC) dimer. The initial expected suppression of conductance was succeeded by a temporary relapse of conductance. Such an anomalous result was explained by the competition between the vertical decoupling of π-states by stretching, and the rotational alignment coupling due to the twist. This work has also demonstrated the importance of correlation between the different degrees of freedom in the molecule. Simple independent phonon models would not be able to predict such an anomalous effect.

3. **Band Structure/Hopping Sites:** Although delocalized band structures do not exist in biomolecules, there can be energy gaps in the different hybridized electronic states in the molecule. Additionally, the spatial extent and energy levels of various hopping sites can affect the efficiency of charge transfer. The ionization potentials for different DNA bases are not equal to each other (G < A < C,T) [55]. Therefore, Guanine seems to be the first candidate for hole-hopping [42]. It has been shown, though, that A-hopping is also possible to facilitate charge transfer between distant Guanine bases [56, 57]. To this end, it is important to note that the ionization potentials for isolated DNA bases can be different from their actual values within the DNA double helix. For example, Cuniberti *et al* [58] has shown that the GC base states can hybridize with those in DNA backbone and this way affect the charge transfer. Additionally, the base ionization energies ought to be influenced by the base stacking, in that formation of higher energy anti-bonding states between stacked bases can affect the hole transfer rates. Finally, the hydrogen bonding between matched bases may result in conformation and spatial extent changes for the conducting states (HUMO and/or LUMO). The spatial overlap is essential in hopping between sites, as described by the simple Mott-Davis relationship for the hopping rate in solid-state physics:

$$R = \omega_{phonon} \left[ e^{-\alpha|r_1 - r_2|} \right]^2 e^{-\Delta E / k_B T}, \tag{3}$$

where $\omega_{phonon}$ is the phonon frequency, $\alpha$ is the distance decay factor, and $r_1$ and $r_2$ are the position vectors for the localized hopping source and target.

Charge hopping through peptides is even more complicated, as there are several different candidates for hopping sites. The peptide bonds have resonance forms which allow bonding electron to be transported through them. Coupled with molecular motion, this can be an efficient method of charge transport. In addition to the peptide bond, certain residues have been shown to facilitate efficient hopping. Shih, *et al* [59] has reported a substantial increase in the hopping hole transfer rate in peptides with abundance of Tryptophan. Aromatic residues are therefore seen as candidates for efficient hopping sites.

A lot of theoretical effort attempted to identify different conductive pathways in biomolecules, or phenomena that lead to the intermittent presence of such pathways. Theoretical work on hopping generally attempts to estimate hopping energies in candidate sites, as well as any corresponding energy gaps within the hopping site, or the environmental effects to create or destroy potential hopping sites. In [60], Apalkov *et al* utilized a Hamiltonian model to describe different types of HOMO-LUMO hopping between adjacent bases. Their model was able to predict a gap of 0.75 eV for A-T pairs and 1.1 eV for G-C pairs. In the case of DNA, there is a general belief that hopping can theoretically occur along the double-helical axis, across the bridge, or diagonally between bases. Hawke *et al* [61] recently applied *ab initio* simulations to resolve the electronic structures of the isolated base pairs and DNA base dimers, resulting in theoretical estimation of hopping parameters that could be used in transport models. Kubar *et al* [62] used a coarse-graining approach to calculate the coupling elements between different bases (necessary for hopping transport calculations) with a high degree of accuracy, while being computationally efficient enough to allow coupled molecular dynamics simulations. The effect of solvation molecules on the creation of hopping sites was also studied theoretically in [63]. It was shown that wet DNA reduces the HOMO-LUMO gap to 3 eV as opposed to 8 eV in the dry state. This suggests that the electronic states and conformations of counter-ions and solvent molecules can affect the hopping rates. It is therefore expected that the surrounding environment's vibrational degrees of freedom will couple to that of the molecule, further complicating the calculation of energy states.

The modelling attempts above, in addition to many others, show that simplified treatments of carrier transport are not sufficient for biomolecular conduction. The hopping and tunnelling rates are very sensitive to the electronic structure, which, in turn, is highly responsive to the surrounding medium and to the dynamics of the molecule. Furthermore, radiation-induced oxidation affects the electronic energies in a manner different from electrode-applied biases. This makes the two different experimental approaches incompatible in determining the carrier transport dynamics. The large perturbations in electronic states and energies necessitate the use of advanced quantum formulations of the transport problem. The Keldysh formalism (also known as the Non-Equilibrium Green's Function of NEGF approach) is often used to describe the carrier dynamics in such a case.

## 6. Critical Assessment of the Biopolymer Charge Transfer/Transport Theories

As we have seen from the above consideration, especially for DNA electrical properties, there is still no general consensus as to the plausible physical-chemical mechanism of the charge propagation. With this in mind, we shall scrutinize the modern theoretical approaches to describe charge transfer/transport through biopolymers and try to show possible ways of theory and model refinement.

Usually, tight-binding Hamiltonians are employed to describe electronic structures of DNA duplexes – both explicitly (in the form of the "fishbone", "ladder" and similar models – see, for example, [51, 53, 58, 60, 64-72] and the pertinent review articles [3, 73-77]) and implicitly (within the framework of Marcus-type theories of charge transfer, for example, [55, 78-81] and the references therein). These works were successful in qualitatively (and sometimes even quantitatively) describing numerous experimental data (see, for example, [42, 56, 57, 82, 83] and the references therein) on transfer of injected single holes (or injected single electrons) through DNA duplexes. Such a kind of "tight-binding philosophy" has been analyzed in detail in the work [84]. This work studies homooligonucleotides with regular sequences $(dA)_n$-$(dT)_n$ and $(dG)_n$-$(dC)_n$ by constructing double-stranded tight-binding Hamiltonian with the sites, each representing the corresponding separate purine and pyrimidine bases. Every base is a carrier of one frontier orbital (HOMO, if hole transport is considered, or LUMO, if excess electron is transferred through DNA), with the corresponding ionization potentials (HOMO energies) or electron affinities (LUMO energies) being the "site energies" in the tight-binding Hamiltonian. The true challenge of charge mobility in such a system begins with the introduction of the separate intra- and interstrand electron couplings into the latter representation (see Scheme 1).

Thus, the nucleotides as the "tight-binding sites" are just reduced here to hydrogen atoms. The physical approximation at this rather low level would be correct, when ionization potentials are as high as (or at least comparable to) that of the H atom (The in vacuo ionization potential of hydrogen atom is well known to be equal to 13.6 eV), or electron affinities are high enough. This is but not the case for nucleic acids, as we shall discuss below.

To this end, the fresh result from J. Barton's group is worth mentioning [45]: DNA-mediated hole and electron transfer can both be triggered consecutively within the same DNA duplex. Barton et al. dub this effect "the ping-pong reaction", which likely involves hole migration primarily through the purine strand with electron migration facilitated by stacked pyrimidines, so that the analogous parameters ought to govern both hole and electron transfer through one and the same DNA base-pair stack.

Hence, to theoretically study the latter "ping-pong reaction", the "one-orbital-per-site" picture seems to be utterly insufficient. But why – and how to improve the model ? Let us consider a simplified model of the DNA duplex electronic structure (see Scheme 2).

Indeed, in the zero approximation, let us agree that each DNA base contains not only one, but both frontier orbitals, HOMO and LUMO. If A and T (or G and C) are situated at the infinite distance from each other, the frontier orbitals of such a complex will be degenerate (by properly choosing the zero of energy, we may render HOMOs or LUMOs degenerate). Now, let us bring the A and T closer together, so that the geometrical prerequisite for forming the Watson-Crick hydrogen bonds are fulfilled. Switching on the latter brings rather strong interaction between the bases, as already mentioned (the energy of the A-T attraction is about –13 kcal/mol (-0.6 eV), whereas that for the G-C pair is about –23 kcal/mol (-1 eV) [85]). Physically, such a strong coupling should *lift* the degeneracy of the frontier orbitals, so that the electronic energy levels of the pyrimidine bases will be lower with respect to those of the purine ones. Theretofore, when looking for the frontier orbitals of the whole H-bonded purine-pyrimidine complex, we immediately recognize that the HOMO of this complex is situated on the purine base, whereas its LUMO – on the pyrimidine base. In the homooligonucleotide duplexes with regular base sequences, this qualitative picture will be preserved, with some possible intrastrand shifting of the frontier orbitals due to stacking of the base pairs (which causes more or less strong overlap of the DNA base $\pi$-orbitals, leading to interactions with lower energies than those of the Watson-Crick H-bonding). The frontier orbital distribution in question is supported by quantum-chemical computations at any level of approximation [86-90]. Most recent resolution of DNA electronic structure using a clever combination of transverse scanning tunnelling spectroscopy with ab initio quantum-chemical calculations [19] is in full agreement with the earlier predictions using other quantum-chemical approximations.

Here we have considered just bases, and the attentive reader will tell us, that the base pairs are actually the hydrophobic core of the DNA duplex structure, which is held together by covalently bonded deoxyribose-phosphodiester backbone. Each phosphodiester bridge of the latter carries a negative charge, so that there ought to be enormous electrostatic repulsion between the opposite backbone strands, which largely compensates the base-base attraction due to the Watson-Crick H-bonding. As a result, the effective base-base attraction would not more be so huge as the in vacuo one, so that the electronic energy levels of the bases may most probably remain quasi-degenerate.

Still, this usual reasoning to advocate the Mehrez-Anantram-like approaches is also insufficient. The point is that, in the realistic systems, the DNA duplex polyanion is every time placed into the counterionic atmosphere to fulfil the principle of electroneutrality. In fact, if we consider *in vitro* DNA systems, the most common counterions are alkali cations. Moreover, DNAs are hugely hygroscopic, so that it is most appropriate to speak of the counterion-hydration shell of DNA duplexes. The structure of this shell is known to mimic the DNA double-helical structure, with counterions and water molecules wrapping the DNA polyanion to form the so-called Manning shell. "Manning shell" physically means that the counterions are massively condensed at some proper distance from the DNA surface, with the water of hydration filling the whole space between the DNA surface and the counterion condensate. Furthermore, this Manning shell is moving as a whole with respect to the DNA polyanion, activating the so-called "DNA acoustic plasmons" [91-96]. Hence, the very presence of the counterion-water shell, as well as the collective dynamics of the latter, should physically guarantee both static and dynamic screening of the phosphodiester electrostatics. The latter, being very effectively screened in such a way, ought to allow sufficiently strong interaction between the H-bonded bases, rendering the physical picture of "DNA frontier orbital separation in space" discussed above (see Scheme 2) a quite realistic one.

Why "interstrand electron coupling" is impossible when the DNA frontier orbitals are separated in space, as discussed above ? Let us consider the most popular hole transfer through DNA duplexes. The hole is most probably formed on one of the purine bases which is the HOMO carrier. As – owing to the Watson-Crick H-bonding – the HOMOs of the pyrimidine bases are shifted much lower in energy than those of the purines, the most probable next address for the hole will be another purine site – and so on. For the hole to jump over to the pyrimidine strand would require overcoming of a significant energy barrier. The analogous, but reversed, reasoning obviously holds for the excess electron transfer via LUMOs.

Bearing all this in mind, which consequences should then the "DNA frontier orbital separation in space" representation have for the physically correct and logically consistent modelling of DNA duplex electric properties ? We may suggest two equivalent tight-binding models, depending on the nature of the problems to solve. First, we may declare nucleotide pairs – and not the separate nucleotides – the tight-binding "sites", so that we immediately arrive at the effective quasi-1D model of the DNA duplex. In this model, we may concentrate on only one frontier orbital per pair (HOMO for the hole – and LUMO for the excess electron – transfer/transport). For the homooligonucleotide duplexes with the regular sequences, such a picture would amount to the effective one-band model. Second, if the whole double-helical structure is necessarily needed, one might opt for the effective two-band model, where the valence band is situated on the purine – and the conduction band is carried by the pyrimidine – strand. This kind of model has already been successfully used to predict mediation of long-range charge transfer by Kondo states in DNA duplexes [97], but this interesting and instructive work has regretfully been lost in the avalanche of the "hydrogen atom" models of the Mehrez-Anantram type.

The next very important point concerns the DNA Hamiltonian parameterization. Most frequently, the tight-binding model "site energies" are represented by *in vacuo* ionization potentials/electron affinities. Such a viewpoint is logically inconsistent. As we have already discussed above, it is clear that placing DNA bases into the polymeric surrounding will significantly change their "site energies" compared to the corresponding *in vacuo* values. The extent of such a change is in principle well known from the literature (see, for example, [98] and the numerous references therein), but is for unclear reasons completely ignored in the DNA charge transfer/transport publications. Specifically, the usual choice for the DNA "site energies" are *in vacuo* ionization potentials of DNA bases, which are of order of 7 – 8 eV, whereas already a mere addition of covalently bonded deoxyribose and phosphate lowers the base ionisation potentials to about one half of their in-vacuo values. Hence the final ionization potential value of the oligomeric DNA duplex should in effect lie between 4 and 5 eV [98], being thus in resonance with the 1st singlet transition in DNA (see thorough discussion of this topic in [88]) – that is, about three times as low as the *in vacuo* ionisation potential of a hydrogen atom.

Therefore, to completely omit LUMO when considering DNA duplexes is for the most problems not the best approximation. Indeed – owing to the resonance between the DNA ionization potential and the 1st singlet transition – there ought to be a non-negligible admixture of the DNA excited state to its ground state, which would entail some additional charge separation (so necessary for the DNA electric properties). Moreover, the degree of this admixture might definitely be regulated by some specific conformational modes of DNA duplexes – via pseudo-Jahn-Teller effects, for example – which is a very interesting topic currently studied in our labs. On the other hand, one may say that the Boltzmann factor with the energies of order of 4 – 5 eV is still negligible enough (like that with the energy 13.6 eV) to ensure the applicability of the one-band models in studying charge transfer/transport through DNA duplexes. We completely agree with this statement, but would like to reiterate that, rigorously, the one-orbital/one-band models are applicable solely to the transfer of an *injected* single hole/electron through DNA hydrophobic core. More complicated processes, like, for example, the "ping-pong reaction" [45], definitely require more elaborate models.

To this end, the very functional form of the conventional tight-binding Hamiltonian may also be insufficient for some problems connected with the DNA duplex electrical/magnetic properties. Indeed, the work [88] has underlined appreciable importance of electron-electron correlations in DNA duplexes (well and long known from the numerous works by Ladik et al., see, for example, [99]), which would require recasting the DNA Hamiltonian as an extended-Hubbard-type one.

Finally, DNA is well known to be essentially a molecular-dynamical system, which enables its representation as a true "soft matter". DNA dynamics is known to dramatically affect the relevant charge transfer/transport processes, as demonstrated by the pertinent systematic experiments (see, for example, [100, 101] and the references therein). Meanwhile, theoretical description of these effects is usually carried out at the level of polaronic theories. Although the polaronic models are capable of fitting the experimental data on injected single hole/electron transfer, such a model of DNA dynamics is really very naïve. Dynamics of biopolymers includes motions of sufficiently high amplitide and extremely broad, quasi-continuum spectrum of characteristic times, with DNA duplexes being no exclusion from this general rule. Such dynamics cannot be described by determinististic equations of motion and requires involvement of intricate non-linear diffusional processes [102]. On the other hand, polarons are local, linear deformations of some rigid lattice in response to the electron/hole arrival at this particular spot in a macromolecule. Such models, although extremely popular up to the most recent time [46], should definitely be inconsistent with the notion of DNA as a typical example of a "soft matter". To this end, the work [103] has systematically investigated interaction of DNA electronic subsystem (considered in the one-band approximation) with the conformationally active vibrational modes and found that tight-binding site energies and hopping integrals (chemically, frontier orbital energies) are in most cases non-linearly dependent on the conformationally important normal modes in DNA duplexes. Such effects are rather commensurable with the "rest-and-fire" charge-transfer mechanism [39] mentioned in the previous section, than with the polaronic picture. According to the work [103], the latter representation can solely be used to describe DNA electron system's interaction with the "strecthing-squeezing" transversal vibrational mode in DNA duplexes. Thus, it is throughout possible that the actual mechanism of charge transfer/transport through DNA duplexes represents a bizarre mixture between polaronic and non-polaronic contributions.

Further important insights into the influence of DNA molecular dynamics on its electric properties can be gained by employing all-atom molecular-dynamics simulations in conjuction with reliable quantum-mechanical methods to describe electronic structures of – and/or processes in – DNA duplexes. For the present, the most frequently used quantum-mechanical approaches in such studies comprise either tight-binding Hamiltonians (with all their merits and demerits discussed above) and DFT (density functional theory) (see, for example, [54, 62, 104-108] and the references therein). Employing the latter also produces many serious problems, like restrictions on the molecular dimensions of model DNA duplexes (still, as mentioned above, the O(n)-linear-scaling DFT approximations might help in partially resolving this poser [105]), as well as the role of atomic basis set quality. A rather poor quality of the atomic orbital basis set may produce strange unphysical results, like positioning of LUMO on $Na^+$ counterions [109], which is obviously a well-studied quantum-chemical artefact [110]. Bearing all this in mind, we are sure that in the future studies the all-atom MD simulations should be combined with the all-atom techniques to directly estimate charge transmission probabilities. Such methods are still in their infancy and rather rarely employed [52, 111-115], but they definitely deserve much more attention of the scientific community.

Next, we would like to present a brief qualitative outline of a new DNA charge transfer/transport theory, the work on which is in progress in our lab.

## 7. DNA Charge Transfer/Transport as a Diffusion-Controlled Reaction

As we have seen above, the polaronic models of DNA charge transfer/transport still widely popular nowadays are physically rather naïve, so that the parameters obtained by fitting such models to the pertinent experimental data (see, for example, [101,102]) are not relevant to the actual DNA molecular dynamics. In our opinion, more appropriate models should and could be suggested: for example, one might directly treat travelling of the charge carriers through DNA duplexes as a kind of diffusion-controlled reaction consisting of sequential charge transfer steps between the neighboring Watson-Crick H-bonded pairs of nucleotides (or even stacks of such pairs). In the conventional terms (see, for example, [56]), every step can be seen as "superexchange" or "hopping", whereas the whole reaction – as a sort of "gradual charge diffusion" (according to what we have discussed above [57,58], the latter could be possible in sufficiently long DNA duplexes only).

A generalized theory of linear chains of diffusion-controlled reactions (with the possibility of branching processes) has been worked out about 20 years ago [116,117], so that we would solely need to correctly apply this approach to the DNA charge transfer/transport.

The model [116,117] considers an overall reaction $\mathbf{R} \rightarrow \mathbf{P}$ ($\mathbf{R}$ = reactant, $\mathbf{P}$ = product) which consists of $i = 1, 2, \ldots, N$ elementary steps (dubbed as "regimes"). For DNA duplexes, a "regime" would physically mean the result of charge transfer between two neighboring DNA duplex units (that is, chemically, cation- or anion-radical state). Such a "unit" could, in principle, consist of one nucleotide pair with the bases forming a (Watson-Crick – or non-Watson-Crick) H-bonded complex – or maybe a nucleotide pair dimer with its Watson-Crick base pairs stacked, depending on the localization degree of the pertinent frontier molecular orbitals.

The theory [116,117] considers elementary processes *within* the "regimes" first. Physically-chemically, in our system, these ought to be based upon the frontier orbitals of the Watson-Crick H-bonded nucleotide pairs. Therefore, the process within the "regime" should be the time evolution of the corresponding cation- or anion-radical states. Further, we need also a clear physical-chemical picture of the elementary transitions between the "regimes". For the DNA duplexes, one might basically envision two types of the latter – the *superexchange* (if the frontier molecular orbitals of the neighboring Watson-Crick H-bonded nucleotide pairs in the stack overlap in the double-helical axis direction – or, in other words, there is a partial orbital delocalization) or the *charge hopping* (when the frontier orbitals of the neighboring Watson-Crick H-bonded nucleotide pairs in the stack do not overlap at all, which means that these orbitals are perfectly localized).

With this in mind, we may wish to define the accessible state space of the "regime" *i* as $X_i$. We assume that, if the charged particle arrives at the regime $X_i$ at some time point *t*, it may stay in this regime for some time – and then leave it in one of the two opposite directions, to escape either to $X_{i+1}$ or to $X_{i-1}$. To be able to enter one of the latter regimes, each of the $X_i$ should be possessed of 4 boundary (entrance and exit) states, because it has two borders and two possible directions for the charge entrance/exit. Besides, we state that every entrance/exit event has some probability of occurrence (for details, confer [116,117]). Finally, the regimes $X_0$ and $X_N$ are considered stable reactant (**R**) and product (**P**) of the total reaction, respectively, so that they have only 2 border states each.

As the above-mentioned state-transition probabilities should clearly be time-dependent, we may write down some equations of motion for them, assuming that the evolution of the probabilities within every regime is independent of each other (that is, the elementary stochastic processes are Markovian, although the total process might even be non-Markovian as well). The main interest here would lie in the long-time limits of the probabilities in question, because the former are directly connected with the rate of the diffusion-controlled reaction under study.

We send the interested reader to the original works [116,117] for the mathematical derivation details. Here we would only like to stress the following. The theory outlined above allows us to estimate the *apparent* rate constant $k_{app}$ of the total diffusion-controlled reaction of the DNA charge transfer/transport type. Specifically, we get by and large to some modification of the fundamental Smoluchowski equation:

$$\frac{1}{k_{app}} = \frac{1}{k_{transp}} + \frac{1}{k_{chem}} \,, \tag{4}$$

Here, the first reciprocal term on the right-hand side of this equation denotes the so-called "transport" rate constant, which is the rate constant of the diffusional/(any complicated non-chemical) process taking part in the reaction, whereas the second one is the "chemical" rate constant of the total chemical reaction proper – just as it would be in the absence of the diffusional process.

Taking all the above into account, we know, on the other hand, that the experimentally observed characteristic times of DNA charge transfer/transport lie in the picosecond range (see [57,58], for example). That gives us the correct order of magnitude of the pertinent apparent rate constant. Further, it is well established that the characteristic times of the typical elementary charge-transfer reactions are in the femtosecond, whereas those of the "biopolymer conformational diffusion" – in the subpicosecond to picosecond range.

Substituting this information into the right-hand side of (Equation 4), we obtain very neat qualitative agreement with the relevant, experimentally measured apparent constants.

Our work on the model outlined above is still in progress, and we shall present a detailed account on this theme elsewhere.

## 8. Examples of How to Use DNA and Protein Electrical Properties in Practice

### 8.1. DNA as Logic Gates

Aside from their intended use as electronic conductors or at least scaffolds for the fabrication of conductors, DNA molecules have been investigated for the direct implementation of logic functions. It is a known result that all combinational logic can be reduced to a two level multi-input AND/OR circuit with possible inversion at the inputs. The work in [118] demonstrates a technique using DNA in which logic gate functions can be realized. The method depends on charge transport along a DNA molecule. The segment is initiated and terminated with triple G bases (GGG), which are known to possibly capture holes and be oxidatively damaged by electron transport. One end of the ssDNA molecule is tagged with an oxidizing agent to introduce a hole into the DNA strand. The capturing base sequence next to the oxidizing agent is termed the "a" capturing segment, whereas that at the end of the ssDNA is the "b" capturing segment. The OR functionality is achieved simply by mixing several different DNA strands that correspond to these different AND operations together. The final output is a (GGG) damage measure from the whole ensemble, which is the ratio of damage to GGGB sites to that of the GGGA sites. This ratio should be close to unity when the DNA strand is conducting, and close to zero when it blocks the hole transport.

The operation of the AND gate is as follows: a logic '1' output is exhibited by passage of the hole to the GGGB sites. In order for this to happen, the ssDNA must conduct current. The authors use the fact that GC pairs conduct current, while other pairs do not. A new purine was constructed (Methoxybenzodeazaadenine MDA) that conducts current when attached to Thymine. Now, both the G-C pair and the MDA-T pair conduct current well. The idea is to put on the tagged ssDNA bases that are either MDA or a G. Now, the input is given by an input ssDNA that has "CCC" bases at its terminals to connect with the GGG strands of the tagged ssDNA. The inner bases of the input ssDNA will have either C base (logic 0) or T (logic 1). The G base in the tagged input is seen as an inverting input, whereas the MDA is a noninverting input. Thus, upon DNA hybridization, we have one of four different possibilities: MDA-C means an input of logic 0, MDA-T means an input logic 1, G-T would correspond to inverted logic 1 (logic 0), and G-C would correspond to an inverted logic 0 (logic 1). Thus, only if the AND functionality is satisfied will there be hole conduction across the hybridized strand be possible. It is important to notice that the pairs MDA-C and G-T are not thermodynamically favourable, and such an orientation would not result in hybridization. However, both combination give a logic output of zero, which is an effective termination of conduction. Thus, hybridization is not even necessary in this case since the output will be zero anyway !

This work was simply a demonstration of the capability of DNA molecules to accomplish logic functions based on their conductivity. However, this is not to say that such techniques will compete (at least not yet) with contemporary microelectronics. Firstly, it takes time for the input to be applied (hybridization).

Secondly, output readout also takes time, since it requires chemical techniques (gel electrophoresis) to assess the amount of radiative damage. Finally, the output measure is a stochastic value that is alarming and can be easily contaminated. However, what is important here is that the "calculation" step is done extremely fast and using an extremely small area. This is very important in extremely large scale integration. In order to fully utilize such a "DNA calculator", one must devise fast and effective means of applying the input and reading the output.

## 8.2. A Protein Diode

Another example could be a rectifying diode made of a single protein [119]. In this work, the blue copper protein azurin (az), which normally includes a doubly oxidized copper ion and which acts as an electron transporter, was immobilized between two gold nanoelectrodes. The authors attempt to explain the rectification by realizing that the folded azurin possesses a permanent dipole moment, and that upon self-assembly on the gold nanoelectrode, the dipole moment of all proteins add up and induce an electric field that has a similar effect to the electric field in the depletion layer of a p-n diode. However, in their work, the authors do not seem to comment on the possibility that one of the electrical contacts to the protein monolayer might have created a Schottky contact, which in itself would be great because it would not require an aggregate effect of many molecules and a molecular Schottky diode would still be quite desirable. In both cases, a nanomolecular diode means that potentially all logic gates could be nanofabricated. This is a truly amazing achievement if it could be reproducibly realized.

## 9. Medical Applications of Single Molecule Conductance

Besides the potential use of biomolecules for nano-scale electronics, the conductive properties of these molecules can also be exploited for designing therapeutic equipment. An example of this is a DNA hybridization sensor. A bridge oligonucleotide between two electrodes can be utilized as a sensor for the complementary DNA strand. The detection is fully electronic as current will be inhibited when there is no hybridization. Furthermore, the highly sensitive dependence of the current on mismatches can be utilized to detect single and multiple nucleotide mismatches. Such a sensor would be a single-molecule detector which circumvents the need for statistical analysis of the hybridization data. It would enable detection of extremely small concentrations of target DNA in a solution, eliminating the need for overcomplicated PCR (Polymerase Chain Reaction) tests, allowing rapid analysis and early pathogen detection. And it is the early detection, as well as the control of outbreaks, that are major cornerstones of the contemporary biosensor research [120-123].

Clinically, single molecule DNA and protein sensors can allow for faster study of the effects of inhibitory drugs, by directly measuring their effect on the conductance current. Drugs might be able to alter native conformations of proteins and possibly interfere with their current-conducting properties. Thus, the sensed current would be an indication of the damage done to the protein/DNA using the treatment. A single molecule sensor could also be used to examine the mechanism of operation of a drug. By identifying the different current-carrying mechanisms/paths within a fully folded protein, it might be possible to identify the exact effect of the drug on the protein/DNA, by examining the amount of current change and correlating it with the several current conducting pathways.

Bearing this in mind, we may say that single molecule detection does away with the usual, bulk experimentation of biochemical experiments and allows a direct, molecular level probing which gives us a great edge in successful synthesis of effective drugs. The physical-chemical modalities of the above-mentioned sensors are hot topic for the present, they are systematically studied by several groups using theoretical approaches and cause vivid debates in the literature (see, for example, works [124-131] and the references therein).

Finally, looking at the developments of the last several months, one might conclude that several noticeable successes have been achieved in the experimental studies of single-molecule DNA charge transfer/transport. In the work [132], for example, stable and reproducible electronic conduction through double stranded (ds) DNA molecules in a nominally dry state was observed. In this case, to get a stable conduction, 15 base-pair

guanine:cytosine rich dsDNA was immobilized between gold nanogap junction using the conventional thiol linkers, with the DNA polyanion stabilized using a polycation spermidine (Sp$^{+3}$). The electrical measurements have been performed both in air and in a nitrogen atmosphere, and it is the latter that enables the stability and reproducibility. The findings of this work are of extreme interest for studying DNA conduction mechanisms, as well as for designing dsDNA-based biosensors [133-134].

Although pure single-stranded (ss) DNAs are well known to be much worse conductors that the dsDNAs (see, for example, [113]), one should not completely discard them as possible versatile components of future molecular wires. To this end, the work [135] has demonstrated that, if one wraps ssDNA around single-walled carbon nanotubes (SWNTs), the semiconducting behavior of latter is drastically changed after ssDNA modification, showing a clear charge-transfer process at room temperature. A further systematic investigation into the temperature dependence of the single-electron tunneling features in this SWNT-ssDNA system has led the authors [135] to a conclusion that it is the ssDNA wrapping around the SWNTs that helps create quantum dots in series.

Still, as we have already mentioned above, the theoretical developments in the field – although being numerous (see, for example, [136-141] and the references therein) – are largely retarding in their quality. Indeed, naive and fatally oversimplified DNA models are used to investigate such a complex molecule as DNA, by assuming the latter is a conventional semiconductor, like silicon or germanium. That such models might sometimes deliver agreement with experimental data is rather coincidential and in no way advances our understanding of the actual physical chemistry/chemical physics behind DNA charge transfer/transport. Meanwhile, one can distinguish the work [137] among the recent ones in the field, because it shows a bright example of careful and thoughtful analysis of what is known in DNA physics and chemistry during several decades. In our view, theoretical work in the field should be concentrated on systematic quantum-chemical studies (see, e.g., Ref. [142,143] and the references therein, with the most recent paper [143] being an example of a careful and thoughtful usage of the powerful quantum chemical machinery). Another important direction of research would be to work out correct physical models describing DNA molecular dynamics, its intrinsic structural metastability (see, for example, [144-148] and the references therein for the general description of the problems) and their role for DNA charge transfer/transport.

## Conclusions

The electrical conductance of DNA and proteins remains a subject of intense research. While many scientists seem to have completely abandoned the use of DNA as molecular wires, favouring carbon nanowires or nanotubes, DNA molecules still possess the strong advantage of self-assembly and extreme specificity. Proteins might not be intended as wires but their electrical conductivities surely play a big role in their function as structural/transport units, or enzymes. In fact, some of their properties might trigger their use (in crystallized form) as new materials for electrodes, electronic devices, and sensors. In this report, several concepts relating to the theory of transport in DNA and molecules were highlighted, from quantum-mechanical concepts to probabilistic treatments. It can be clearly seen that different modeling domains are quite active in research and that they all show some level of explaining and even predicting experimental results. On the other hand, successful utilization of these conductivity theories requires a high degree of control in design and results that are quite reproducible. Additionally, simple mathematical formulations of conductivity are needed, if these devices are going to make it to commercial compact-model simulations for nanoelectronics. Ab-initio simulations are in fact not feasible when dealing with nanoelectronic circuits, so that simpler adequate and accurate models must be utilized. Transport theories working well for inorganic and organic semiconductors are not immediately adaptable to biological molecules. However, they do provide some hints to possible electron transfer theories and some of the resulting equations can be crudely adapted, at least as a first step towards building the proper full-fledged models.

**Acknowledgements**

The authors acknowledge support from the Canada Research Chair program and the Natural Sciences and Engineering Research Council (NSERC) of Canada, the WCU (World Class University) program sponsored by the South Korean Ministry of Education, Science, and Technology Program – Project no. R31-2008-000-10100-0 – and POSTECH, for supporting this work. This work has also been supported by the Deutsche Forschungsgemeinschaft (DFG) within the Priority Program 1243 "Quantum transport at the molecular scale" under contract CU 44/5-2, and by the European Union under contract IST-029192-2 "DNA-based nanoelectronic devices". We further acknowledge the Center for Information Services and High Performance Computing (ZIH) at the Dresden University of Technology for computational resources. Finally, the authors gratefully acknowledge the valuable comments and suggestions from the anonymous reviewers.

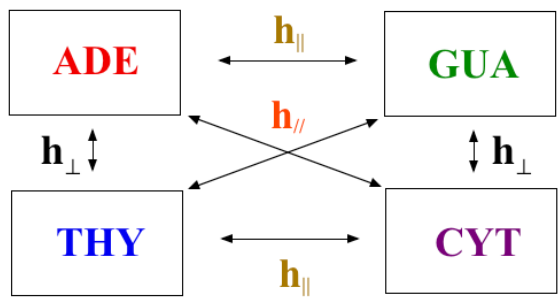

**One-orbital representation
of DNA duplexes**

**Scheme 1.** (Compartments denote a DNA base with only one orbital. The color shows that the site energy (ionization potential) for each base is different. Arrows demonstrate possible electron coupling matrix elements (hopping integrals, "h") of this tight-binding model.)

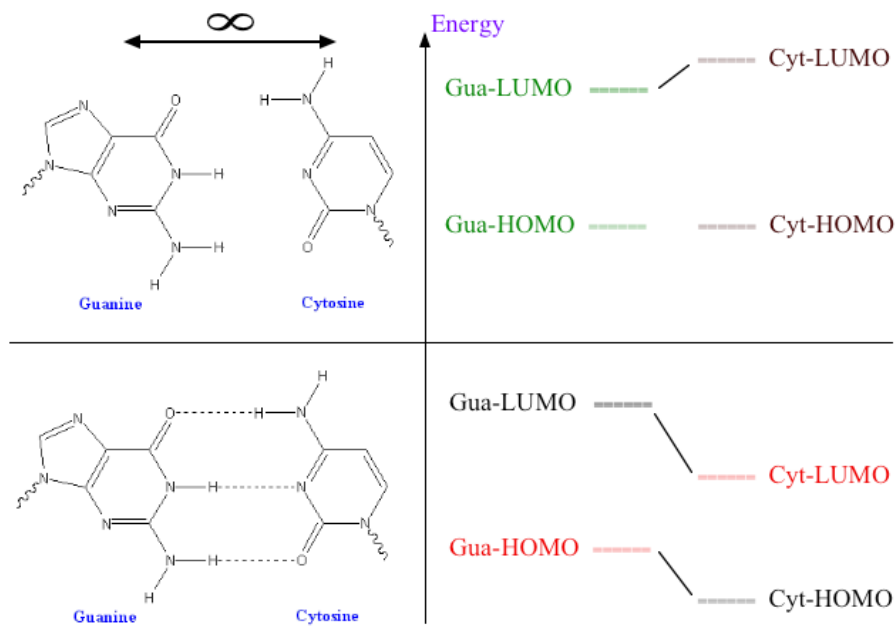

**Scheme 2.** (Two-orbital model of DNA base pairs (here, on the example of guanine-cytosine pair), with the mechanism of frontier orbital separation due to Watson-Crick hydrogen bonding, with the guanine HOMO being the HOMO of the Watson-Crick complex and the cytosine LUMO – the LUMO of this complex)

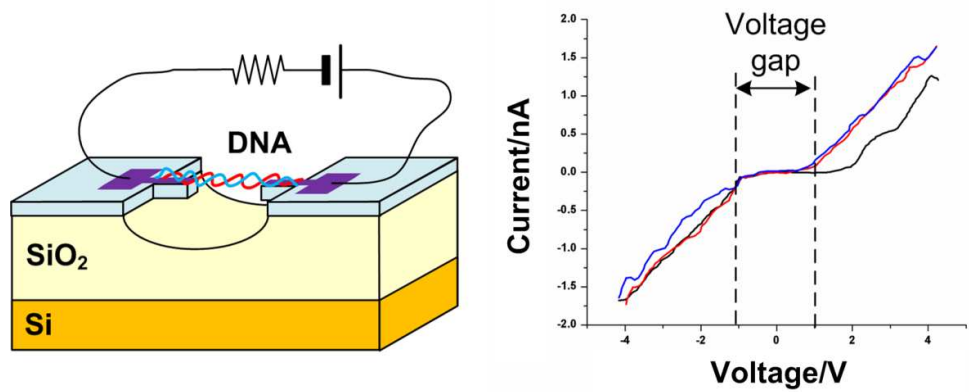

**Figure 1.** (Illustration of Porath's experiment and resulting I-V curves. Reproduced by permission from Macmillan Publishers Ltd: Nature [8], copyright 2000)

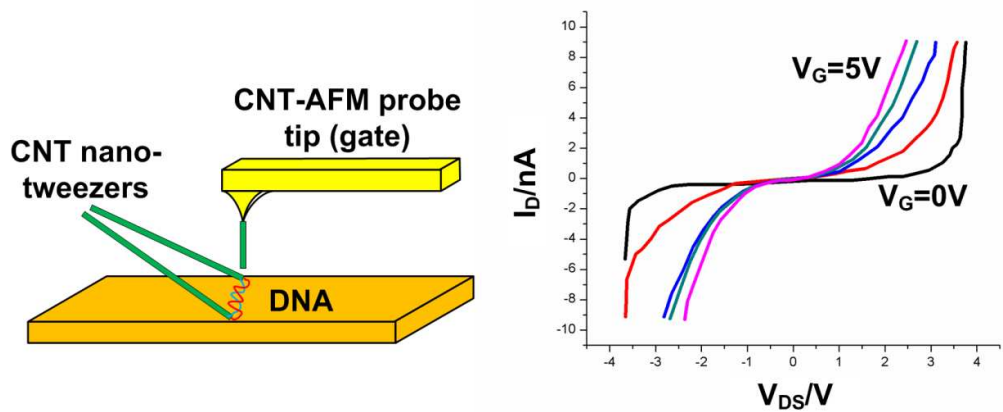

**Figure 2.** (Watanabe's nanotweezer-AFM probe experiment with resulting I-V curves at different bias voltages, showing modulation of the voltage gap. Reproduced with permission from [10]. Copyright 2001, American Institute of Physics)

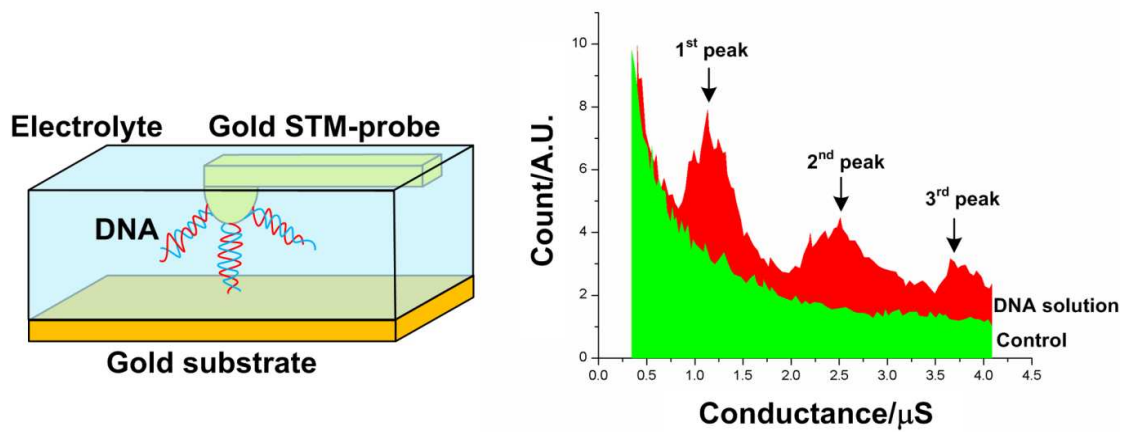

**Figure 3.** (Conductance histogram of a STM-DNA-substrate setup. The peaks show evidence of integer numbers of DNA bridge formation. Reproduced in part with permission from [11]. Copyright 2004 American Chemical Society)

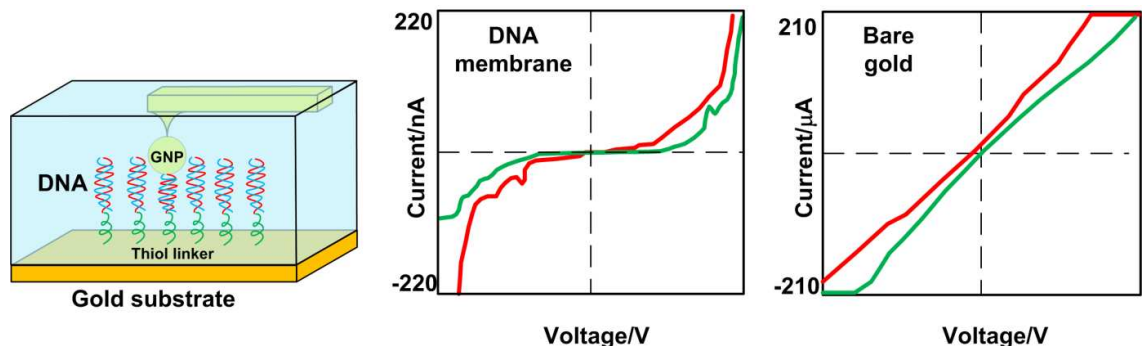

**Figure 4.** (Cohen's DNA gold nanoparticle experiment. Typical voltage gap I-V response was obtained with a DNA membrane, whereas Ohmic response was seen with the AFM probe on bare gold. The green and red curves are forward/reverse runs, and the hysteresis effect was interpreted as evidence of good contact formation [13] (permission requested))

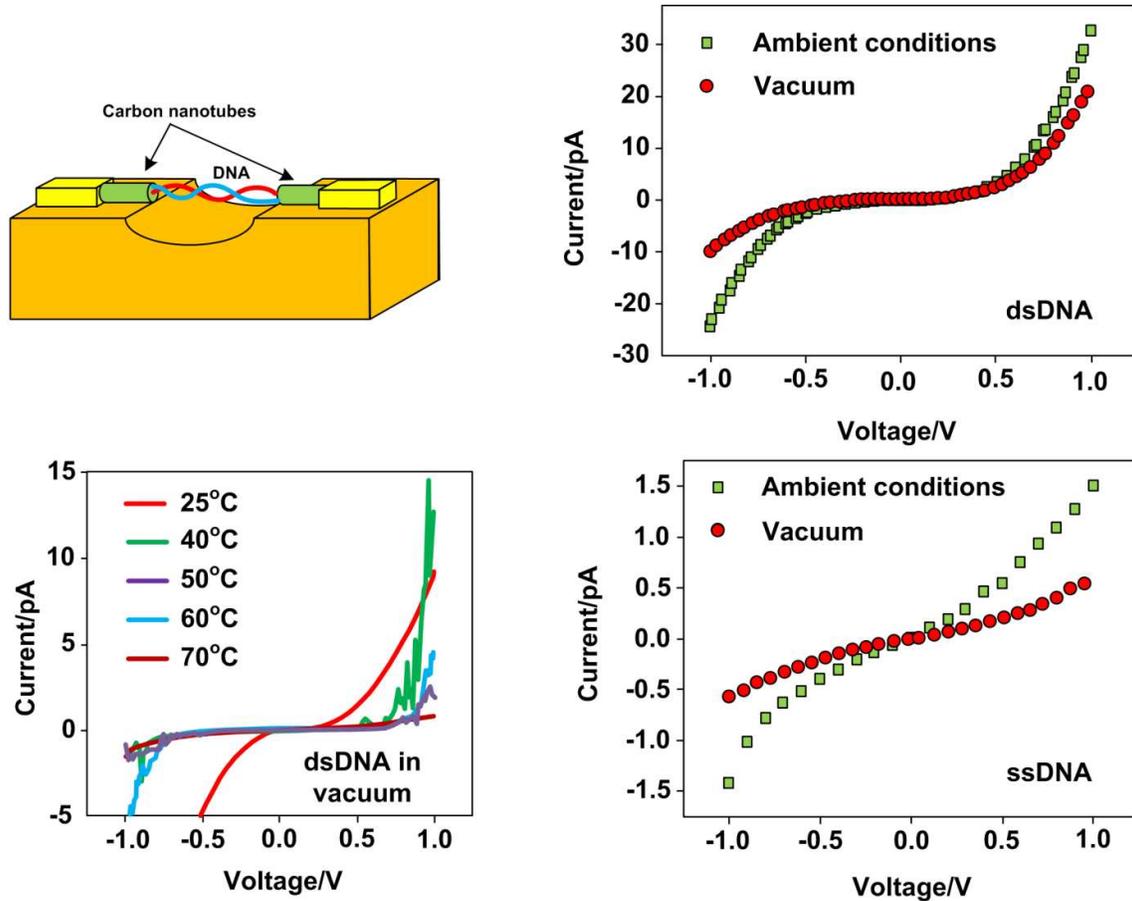

**Figure 5.** (Illustration of Roy's experiment. Results show a comparison of the conductance of suspended double- and single- stranded DNA tethered to carbon nanotubes. Also shown is the variation of the voltage gap with increasing temperature, upto higher temperatures where conductance is inhibited due to denaturation. Reproduced with permission from [15]. Copyright 2008 American Chemical Society)

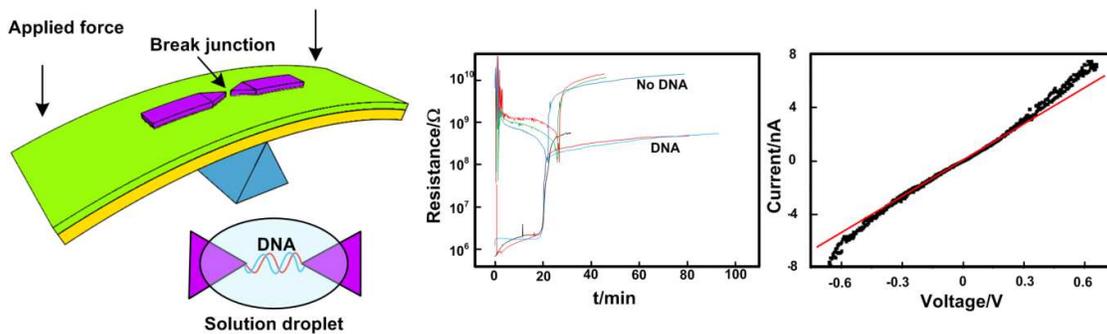

**Figure 6.** (Illustration of the MCBJ experiment by Kang, et al. Two orders of magnitude change in the steady state resistance is attributed to successful DNA junction formation. The molecular junction I-V characteristics shows deviation from the linear relationship typical in tunnel junctions at low bias [18] (reproduced with permission from [18]. Copyright IOP Publishing Limited)

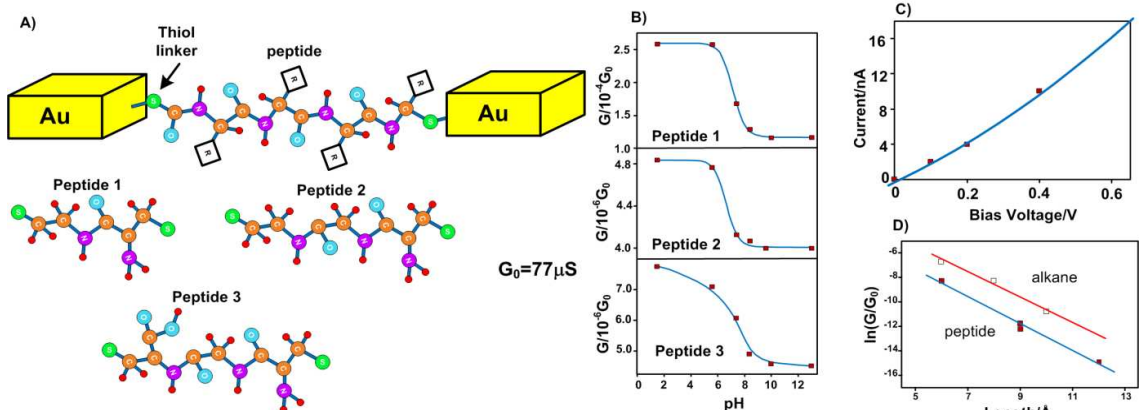

**Figure 7.** (The experiment of Xu et al. A) Experimental setup: synthetic thiolated peptide bridging a gap between two Au electrodes. Also shown are the three different peptides used. B) decrease of conductance with pH increase in all three model peptides. C) Sample I-V curve, extracted from conductance histogram data. D) Comparison of conductance dependence on molecule's length in peptides, as compared to that of a model alkane. Reproduced with permission from [28]. Copyright 2003 American Chemical Society)

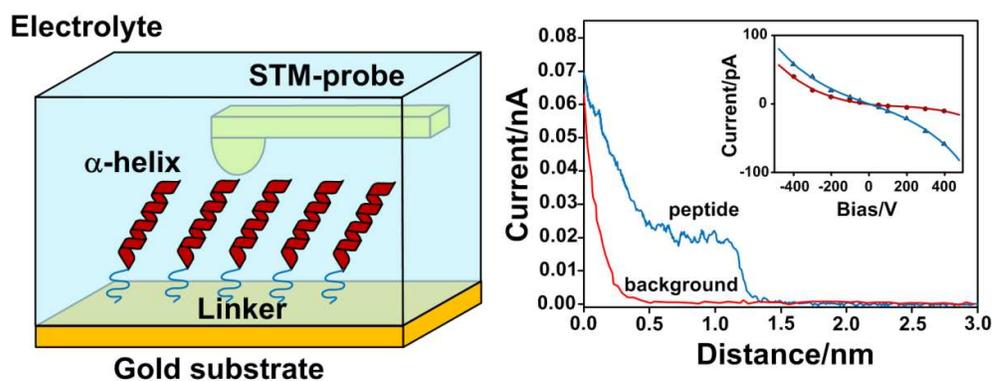

**Figure 8.** (Illustration of Sek's experiment. Left: STM probe contacting a monolayer of α-helix peptides. Right: Current conduction as a function of the tip-substrate distance. It can be seen that the peptide membrane exhibits considerable conductance over that of a bare gold surface. Inset: Sample I-V curves taken at a fixed tip distance for thiol linkers (circles) and peptides (triangles), showing a significant increase in peptide conduction current at positive biases. Reproduced in part with permission from [29]. Copyright 2006 American Chemical Society)

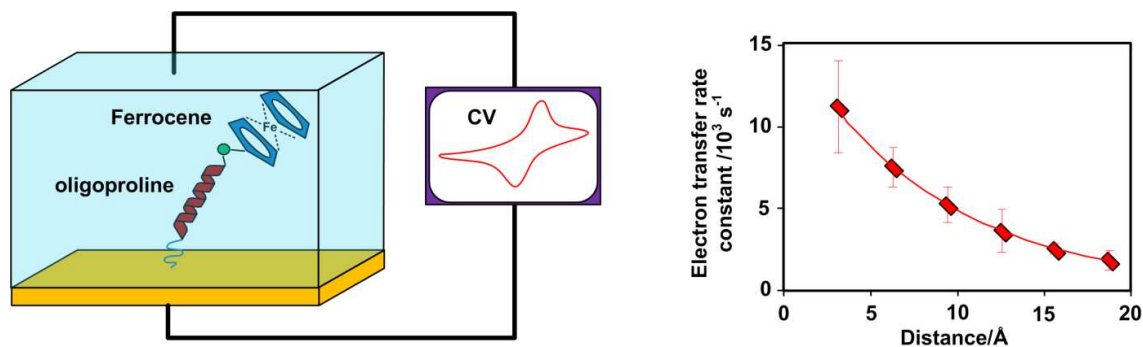

**Figure 9.** (Measuring the conductance properties of oligoprolines using redox-active ferrocene and cyclic voltammetry [30]. The experimental curve shows the exponential reduction of the electron transfer rate with increaing length of oligoproline.)

**Table 1.** (Reported activation energy for various proteins. Reprinted in part with permission from [25]. Copyright 1969, American Institute of Physics)

| Protein | Activation energy [eV] |
|---|---|
| Hemoglobin (methanol denatured) | 2.75 |
| Hemoglobin (Native gel) | 2.8 |
| Hemoglobin (crystalline) | 2.3-2.4 |
| Methemoglobin (crystalline) | 2.5 |
| Serum albumen | 2 |
| Cytochrome C (amorphous, oxidized) | 2.2(70-100$^{\circ}$C) 1.3(40-70$^{\circ}$C) |
| Wool | 2.2 |
| Chloroplasts | 2.1 |
| Rods | 2.3 |
| Gelatin | 2.2 |
| Chlorophyll-protein complex | 2.3 |
| Rhodopsin | 2.3 |